\documentclass[reprint,nofootinbib,amsmath,amssymb,aps]{revtex4-2}
\usepackage{orcidlink}
\usepackage{graphicx}
\usepackage{dcolumn}
\usepackage{bm}
\usepackage{xcolor}
\usepackage{array}\newcolumntype{M}[1]{>{\centering\arraybackslash}m{#1}}
\usepackage{soul}
\usepackage{hyperref}

\begin{document}
\preprint{APS/123-QED}

\title{Search for hyperbolic encounters of compact objects in the third LIGO-Virgo-KAGRA observing run}

\author{Sophie Bini $^{1,2}$\orcidlink{0000-0002-0267-3562}, 
Shubhanshu Tiwari$^{3}$ \orcidlink{0000-0003-1611-6625}, Yumeng Xu $^{3}$ \orcidlink{0000-0001-8697-3505}, Leigh Smith $^{4}$ \orcidlink{0000-0002-3035-0947}, Michael Ebersold $^5$ \orcidlink{0000-0003-4631-1771}, Giacomo Principe$^{6,7}$ \orcidlink{0000-0003-0406-7387}, Maria Haney$^8$, Philippe Jetzer $^3$ \orcidlink{0000-0001-6754-0296}, Giovanni A. Prodi $^{1,2}$ \orcidlink{0000-0001-5256-915X}}
 \email{sophie.bini@unitn.it}
\affiliation{$^1$ Universit\`a di Trento, Dipartimento di Fisica, I-38123 Povo, Trento, Italy}
\affiliation{$^2$ INFN, Trento Institute for Fundamental Physics and Applications, I-38123 Povo, Trento, Italy}
\affiliation{$^3$ Physik-Institut, University of Zurich, Winterthurerstrasse 190, 8057 Zurich, Switzerland}
\affiliation{$^4$ SUPA, School of Physics \& Astronomy, University of Glasgow, Glasgow G12 8QQ, United Kingdom}
\affiliation{$^5$ Laboratoire d'Annecy de Physique des Particules, CNRS, 9 Chemin de Bellevue, 74941 Annecy, France}
\affiliation{$^6$Dipartimento di Fisica, Universit\'a di Trieste, I-34127 Trieste, Italy}
\affiliation{$^7$Istituto Nazionale di Fisica Nucleare, Sezione di Trieste, I-34127 Trieste, Italy}
\affiliation{$^8$Nikhef, Science Park 105, 1098 XG Amsterdam, The Netherlands}
\date{\today}
\begin{abstract}
Gravitational-wave (GW) observations provide unique information about compact objects. As detectors sensitivity increases, new astrophysical sources of GW could emerge. Close hyperbolic encounters are one such source class: scattering of stellar mass compact objects is expected to manifest as GW burst signals in the frequency band of current detectors. We present the search for GW from hyperbolic encounters in the second half of the third Advanced LIGO-Virgo observing run (O3b). We perform a model-informed search with machine-learning enhanced Coherent WaveBurst algorithm. No significant event has been identified in addition to known detections of compact binary coalescences. 
We inject in the O3b data non-spinning third Post-Newtonian order accurate hyperbolic encounter model with component masses between [2, 100] $M_{\odot}$, impact parameter in [60, 100] ${GM}/{c^2}$ and eccentricity in [1.05, 1.6]. We further discuss the properties of the simulation recovered. For the first time, we report the sensitivity volume achieved for such sources, which for O3b data reaches up to 3.9$\pm 1.4 \times 10^5$ Mpc$^3$year for compact objects with masses between [20, 40] $M_{\odot}$, corresponding to a rate density upper limit of 0.589$\pm$0.094 $\times10^{-5}$Mpc$^{-3}$year$^{-1}$. Finally, we present projected sensitive volume  for the next observing runs of current detectors, namely O4 and O5.
\end{abstract}
\maketitle

\section{\label{sec:intro}Introduction}

Gravitational-wave (GW) observations include to date about 90 events, originating from the coalescence of compact binaries (CBC) \cite{gwtc1, gwtc2, gwtc3, nitz20234, mehta2023new}. Such detections provide unique information about compact objects, such as the distributions of their physical parameters and the merger rates of binary systems \cite{abbott2023population}. As detector sensitivity increases \cite{abbott2020prospects}, sources other than CBCs might be detected \cite{O3allskyshort,ligo2021all} and previously unseen astrophysical populations could emerge. 
One such category of sources that could be observed through gravitational wave (GW) emission are close hyperbolic encounters (HE) \cite{capozziello2008gravitational, DeVittori:2012da, bini2021frequency, Grobner:2020fnb}. Dense stellar clusters, like galactic nuclei and globular clusters, are expected to host high number density of black holes (BH) and neutron stars (NS) and can host such scattering events, see e.g. \cite{Gondan:2020svr,rasskazov2019rate}.
During a hyperbolic encounter the bulk of the GW energy is released at the periastron distance. The GW signal manifests as a burst of energy. When hyperbolic encounters are involving stellar-mass compact objects the frequency of the bursts will be in the sensitivity band of the current generation of detectors \cite{dandapat2023gravitational}. 
Apart from the potential HE in the dense stellar environments, HE  can also be of great interest to investigate primordial BH population  \cite{garcia2018gravitational}. Additionally, such interactions are also expected to induce spins in BHs experiencing HE \cite{jaraba2021black}.
The modelling and interpretation of scattering events (also leading to capture) is an active field of research. There are several methods to model HE such as the post-Newtonian (PN) approximation \cite{damour1985general, cho2018gravitational}, post-Minkowskian expansion \cite{khalil2022energetics, damour2023strong}, Numerical Relativity (NR) \cite{jaraba2021black, nelson2019induced} and  effective one body (EOB) formalism \cite{nagar2021effective,albanesi2022new,andrade2023towards}. The recent GW event GW190521 \cite{gwtc3} had an interpretation as a capture event, where the hyperbolic orbits lead to direct capture of the participating BH \cite{gamba2023gw190521}. Also a recent study was conducted to estimate how the capture events can be degenerate with high mass events \cite{Guo:2022ehk}. Recently, a 3 PN-accurate approximant has been used to investigate detection prospects for ground-based detectors  \cite{dandapat2023gravitational}.  

Here, we present the search for GWs from hyperbolic encounters in the second half of the third LIGO-Virgo-KAGRA (LVK) observing run (O3b). We employ the weakly-modeled algorithm Coherent WaveBurst \cite{Klimenko:2015ypf, drago2021coherent}, which is widely used for the detection of generic GW bursts \cite{O3allskyshort, ligo2021all,  szczepanczyk2022all}, CBC signals \cite{mishra2022search,szczepanczyk2021observing} and reconstruction of GW waveform with minimal assumptions \cite{PhysRevD.100.042003}. To enhance the sensitivity for the HE signal, we perform a model-informed search using a decision tree classifier \cite{mishra2021optimization} trained on hyperbolic simulations. To our knowledge, the only other search for HE was proposed by Ref. \cite{morras2022search} which analysed 15 days during the second observing run.

In addition, for the first time, we provide an assessment of the sensitivity volume for such sources, and its constraints in terms of the rate density.

\section{Hyperbolic encounters waveform family}
\label{sec:waveform}
The orbital dynamics of compact binaries on Keplerian orbits influenced by the emission of GWs can be accurately described by the PN approximation to general relativity \cite{blanchet2013gravitational}. In this approximation for slow-moving objects and weak gravitational fields, the corrections to the Newtonian equations of motion are provided in powers of $(v/c)^2 \sim G M / (c^2\, r)$, where $v$, $M$ and $r$ correspond to the relative velocity, total mass and separation of the binary. The advantage of this approach is that one can obtain a Keplerian-type parametric solution to PN-accurate orbital dynamics of compact binaries for general orbits \cite{memmesheimer2004third,boetzel2017solving}. For eccentric and hyperbolic 1 PN-accurate orbits, this was first demonstrated in Ref. \cite{damour1985general}. Once the PN-accurate orbital dynamics are known, one can compute the waveform polarization states $h_+(t)$ and $h_\times (t)$, which describe the GWs emitted by the binary. For hyperbolic orbits this procedure is described in Ref. \cite{deVittori2014gravitational} and in Ref. \cite{cho2018gravitational} for 3.5 PN-accurate orbital motion.

For this work we use the waveform templates provided in Ref. \cite{cho2018gravitational}. Here we give only a brief description of the waveform, for a more detailed explanation of the implementation we refer to Ref. \cite{dandapat2023gravitational}. The quadrupolar order GW polarization states associated with non-spinning compact binaries moving in general orbits can be given in terms of radial and angular coordinates as well as their time derivatives. By adapting the 3 PN-accurate quasi-Keplerian parametrization for compact binaries in PN-accurate hyperbolic orbits derived in Ref. \cite{cho2018gravitational}, we can model the temporal evolution of these dynamical variables for the case of a hyperbolic encounter in terms of a dimensionless PN-parameter $x = (G M n / c^3)^{2/3}$ (where $n$ is the mean motion parameter defined by $n = 2\pi / P$, $P$ being the orbital period which is analytical continuation from the eccentric case), eccentricity $e$ and eccentric anomaly $u$. Due to radiation-reaction, $x$ and $e$ are not static quantities but change at the 2.5 PN order in time, expressions for their time derivatives can be derived from the far-zone energy and angular momentum flux \cite{blanchet1989higher}. 
This yields a system of three coupled differential equations for $x$, $e$ and $u$. With appropriate initial conditions, this system is integrated numerically and the resulting time-dependent functions are invoked into the dynamical variables. It should be mentioned that instead of directly choosing an initial value for $x$, it is more convenient to set the impact parameter when characterizing hyperbolic encounters. The PN-accurate impact parameter $b$ is defined such that $b v_\infty = |\vec r \times \vec v|$ when $|\vec r| \rightarrow \infty$ \cite{blanchet1989higher}, where $v_\infty$ stands for the relative velocity at infinity, and is provided in terms of $x$ and $e$ in Ref. \cite{dandapat2023gravitational}.

It should be noted that, since our waveform family is PN and strictly applies to hyperbolic orbits the approximation breaks down if the black holes come too close or even undergo GW capture. To find the point at which the PN waveform still accurately describes the orbit would involve detailed comparisons with hyperbolic NR waveforms. As these are scarce, Ref. \cite{dandapat2023gravitational} argues that as long as the orbital separation is above $10 \, GM/c^2$, the PN-approximation should be accurate enough. This translates into a restriction on the eccentricity and impact parameter: the eccentricity should not be below 1.15 when the impact parameter is $60 GM/c^2$. For this work we relax the condition on the eccentricity slightly and choose 1.05 as a lower bound. Although the waveform might lack some accuracy, this will not affect the search presented here. Also, we have ensured that a capture for such a system is always excluded since PN formalism breaks down in that regime. Moreover, a search for the capture events was already done in \cite{Ebersold:2022zvz}. 

It should be mentioned that there exist several sophisticated means of modelling GW from hyperbolic events. Currently, the method at the forefront is the EOB approach which can include very high velocities and also the scenario of captures \cite{Nagar:2020xsk,Albanesi:2022ywx,nagar2021effective,albanesi2022new,andrade2023towards}. These models in the future will provide a suitable mean for searches and interpretation of HE also including captures, but they require a larger computational time, 
hence a comprehensive injection study is not possible. We also note that these models are more accurate, but the search presented in this work uses a weakly-modeled algorithm (Section \ref{sec:cwb}), hence the waveforms accuracy and the assessment of their systematics are not crucial requirements, as it is for matched filter searches.

In the following sections, we make use of PN formalism based HE waveform to build the search algorithm and to estimate the search sensitivity for this source. We inject HE with component masses in the range [2, 100] $M_{\odot}$, uniformly distributed in each of the 6 bins $([2,5],[5,20],[20,40],[40,60],[60,80],[80,100]) M_{\odot}$, impact parameter uniform in [60, 100] ${GM}/{c^2}$ and eccentricity uniform in [1.05, 1.6]. We perform about 20,000 injections per mass bin leading to overall 120,000 injections. 

With the aim of finding HE signals in the LVK data we discuss the typical morphology of such signals in terms of their duration and frequency. Generally, heavier masses and larger impact parameters result in longer signals. In this study we consider HE sources with stellar masses which emit a GW signal that lasts less than a second. In the frequency domain, the signal is broadband. The peak frequency increases for lower masses and lower impact parameters for a fixed eccentricity ranging up to $\sim 450$Hz for neutron star masses and down to the low frequency band of current GW detector for more massive systems (Fig. 1 and 4 in Ref. \cite{dandapat2023gravitational}). 
\section{Search algorithm}
\label{sec:cwb}
To search for HE signals in GW data, we employ the weakly-modeled algorithm CoherentWaveBurst (cWB) \cite{drago2021coherent, Klimenko:2015ypf}. 
cWB combines coherently the detector network time-frequency maps, computed with multi-resolution Wilson-Daubechies wavelets \cite{wdm}, and maximizes a likelihood ratio statistic over all sky directions \footnote{The algorithm used here, w.r.t. to the version employed in LVK search for generic short-duration GW in O3 \cite{O3allskyshort}, has relaxed internal thresholds for the selection of the time-frequency pixels containing an excess of power, allowing for the detection of signals with lower signal-to-noise ratio.}. The frequency range analysed is between 16 Hz and 1280 Hz. Each recovered \textit{trigger}, which might be an astrophysical signal or a transient noise, is characterized by several \textit{summary statistics}, as the coherence across the detector network, the peak frequency and the duration.

To assess the significance of each trigger, cWB computes the \textit{background} events distribution from time-shifted detector data. The coincident triggers detected in this way do not have an astrophysical origin by construction, but are due to non-stationary detector noise. The significance is expressed in terms of inverse false alarm rate (iFAR): the false alarm rate of an event with ranking statistic $\rho_i$ is $\mathrm{FAR} = N(\rho_i) / T_{\mathrm{BKG}}$, where $N$ is the number of background triggers detected with $\rho>\rho_i$ and $T_{\mathrm{BKG}}$ is the total background time.

Next, we discuss the capability of cWB to reconstruct HE signals, which qualifies the algorithm for such searches, and we describe the model-informed configuration designed to target HE.
 
\begin{figure*}[!t]
\begin{minipage}[b]{0.3\linewidth}
\centering
\includegraphics[width=0.92\textwidth]{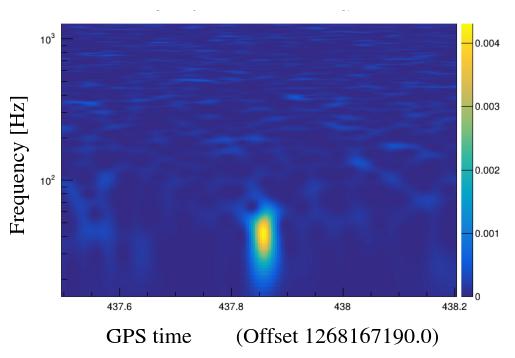}
\end{minipage}
\begin{minipage}[b]{0.3\linewidth}
\centering
\includegraphics[width=0.9\textwidth]{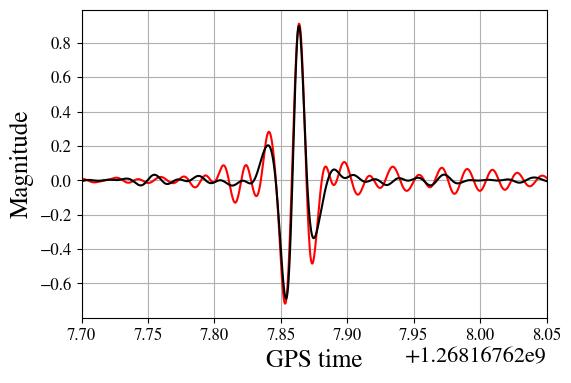}
\end{minipage}
\begin{minipage}[b]{0.35\linewidth}
\centering
\includegraphics[width=0.92\textwidth]{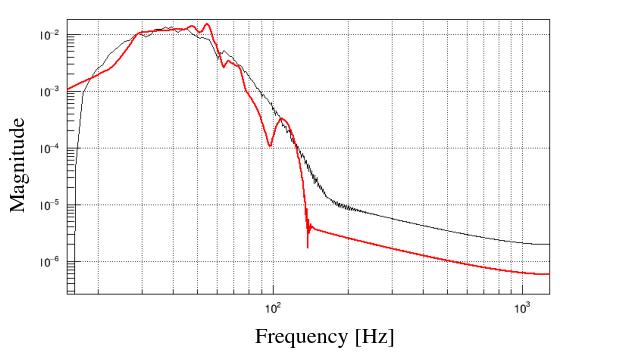}
\end{minipage}
\begin{minipage}[b]{0.3\linewidth}
\centering
\includegraphics[width=0.92\textwidth]{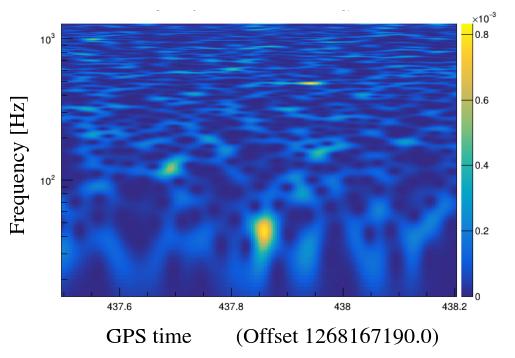}
\end{minipage}
\begin{minipage}[b]{0.3\linewidth}
\centering
\includegraphics[width=0.9\textwidth]{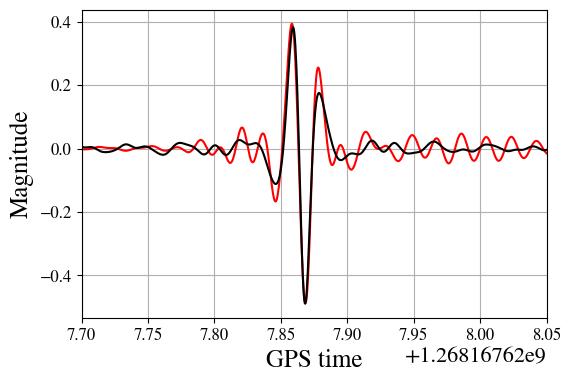}
\end{minipage}
\begin{minipage}[b]{0.35\linewidth}
\centering
\includegraphics[width=0.92\textwidth]{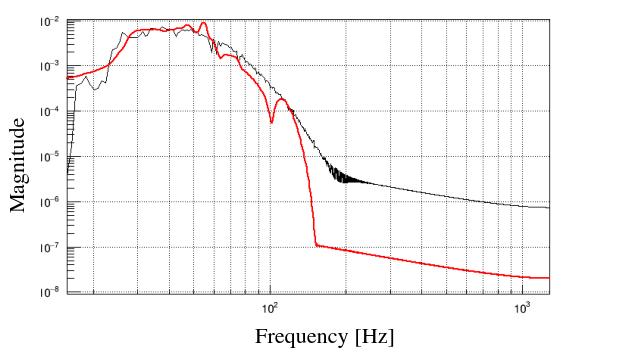}
\end{minipage}
\caption{Example of a simulated hyperbolic encounter event as seen in cWB in LIGO Livingston (top) and LIGO Hanford (bottom). The first column shows the event spectrograms, the second and third columns report the injected waveform (black) and the corresponding cWB reconstruction (red) in time domain and in frequency domain. The approximant is injected at a distance of 182 Mpc with masses  $m_1 = 26.7 M_{\odot}$ and $m_2 = 36.4 M_{\odot}$, impact parameter 59.5 ${GM}/{c^2}$, eccentricity 1.1. The event is recovered by cWB with SNR = 14 and fitting factor (defined in Eq. \ref{eq:ff}) of 0.94.}
\label{fig:encounter_example}
\end{figure*}
\subsection{Goodness of cWB waveform reconstruction}
\begin{figure}[ht]
\includegraphics[scale=0.55]{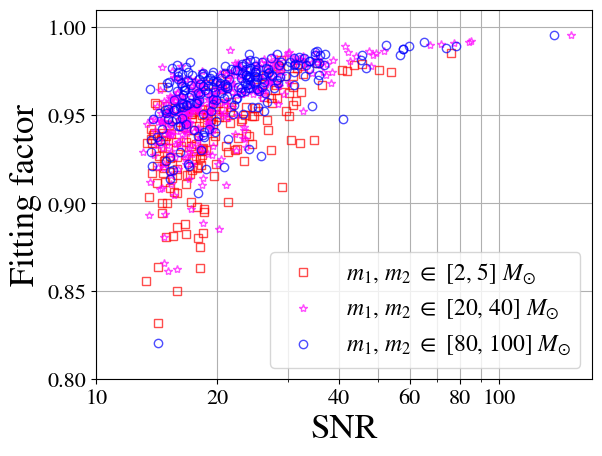}	\caption{Fitting factor versus cWB reconstructed SNR for HE events detected with iFAR $>$ 10 years. HE signals have component masses between [2, 5] $M_{\odot}$ (pink), [20, 40] $M_{\odot}$ (red), [80, 100] $M_{\odot}$ (blue). The FF distributions are peaked above 0.9 meaning that cWB reconstruction is accurate.}
	\label{fig:wav_rec}
\end{figure}

To ensure that cWB is a suitable algorithm for the proposed search, we evaluate cWB effectiveness in recovering HE signals. We discuss the goodness of the cWB waveform reconstruction by injecting HE signals in GW data, and computing the fitting factor (FF) between the injected waveform  $x_{\mathrm{inj}}$ and the one
reconstructed by cWB $x_{\mathrm{rec}}$:
\begin{equation}
    \mathrm{FF}(x_{\mathrm{rec}},x_{\mathrm{inj}}) = \frac{(x_{\mathrm{rec}}| x_{\mathrm{inj}})}{\sqrt{(x_{\mathrm{rec}}|x_{\mathrm{rec}})(x_{\mathrm{inj}}|x_{\mathrm{inj}})}}
    \label{eq:ff}
\end{equation}
where $(x_1|x_2)=\int{x_1(t) x_2(t) dt}$. If the two waveforms are identical FF = 1.
An example of a HE simulation in O3b data as seen by cWB is reported in Fig. \ref{fig:encounter_example}. The event appears as a \textit{blob} at $\sim 40$Hz, and the waveform is reconstructed accurately by cWB.
Fig. \ref{fig:wav_rec} shows the FF versus the signal-to-noise ratio (SNR) reconstructed by cWB for a subset of simulations of HE with component masses in three ranges. As expected, the cWB waveform reconstruction is more accurate for higher SNR signals. The mean values of FF distributions are $0.938 \pm 0.027$ for lower masses ([2, 5] $M_{\odot}$), $0.958 \pm 0.022$ for intermediate masses ([20, 40] $M_{\odot}$) and $ 0.966 \pm 0.016$ for higher masses ([80, 100] $M_{\odot}$), meaning that cWB reconstruction is robust and the injected signal is recovered with high fidelity. It should also be noted that the waveform eccentricity does not affect the cWB reconstruction, as expected.

\subsection{Machine learning signal-noise separation}
CWB algorithm produces a list of triggers, which includes potential GW events and several short-duration disturbances, referred to as \textit{glitches}.  
To reduce the false alarm rate, cWB performs a signal-noise classification based on the decision-tree learning algorithm XGBoost \cite{chen2015xgboost, mishra2021optimization}. XGBoost performs a binary classification between GW signal and noise, learning the characteristics that differentiate the two populations from cWB summary statistics. The noise population is learned from the background distribution, built time-shifting detector data, while the GW signal population is chosen according to the search performed. The output of XGBoost is a number ranging from 0 (for noise) and 1 (for signal) that reweight the cWB ranking statistic.

Here, we train the XGBoost model with a subset (25\%) of the HE simulations (Section \ref{sec:waveform}). We refer to such configuration as a \textit{model-informed} search: it can be considered a middle way between a generic bursts search, which does not make assumptions on the signal characteristics remaining sensitive to a wide range of morphologies \cite{mishra2022search}, and a matched filter search, that instead looks for signals similar to waveform models contained in extensive template banks. Examples of a cWB model-informed search targeting BBH mergers and eccentric CBC are described in Ref. \cite{mishra2022search, ebbh}. Including HE simulations in the XGBoost training dataset, we achieve a consistent improvement in the search sensitivity w.r.t to the generic search. The efficiency at iFAR$>$ 10 years using the HE model-informed search and the generic all-sky one \cite{szczepanczyk2022all} increases from 9\% to 15\% for component masses in the range [2, 5] $M_{\odot}$, from 11\% to 18\% in [5, 20] $M_{\odot}$, from 17\% to 23\%  in [20, 40] $M_{\odot}$, from 18\% to 24\% in [40, 60] $M_{\odot}$, from 16\% to 22\% in [60, 80] $M_{\odot}$ and from 14\% to 20\% in [80, 100] $M_{\odot}$, respectively. The cWB summary statistics used to build the XGBoost model are 10 (the complete list is reported in note \footnote{The cWB summary statistics used to build the XGBoost model are 10: the total energy over all frequency resolutions, cross-correlation coefficient, quality of event reconstruction defined as the residual noise energy over the number of independent wavelets describing the event, square of SNR over likelihood, incoherent energy over likelihood, noise associated to each pixel, central frequency, effective correlated SNR, shape parameter $Q_p$ \cite{bini2023autoencoder}, similarity score to blip glitches \cite{bini2023autoencoder}. Moreover, for this search we do not apply the final correction described in Ref.  \cite{szczepanczyk2022all}, Appendix A.}). Among the others, we include a statistics that measures how much a tested waveform is similar to blip glitches. Such a similarity score is computed by an autoencoder neural network, trained on real glitches occurred in O3b. This method shows to be effective in increasing the search sensitivity, especially for signals with a morphology similar to such transient noises \cite{bini2023autoencoder}. 
\section{Data}
We search for GW from hyperbolic encounters in the second half of the third LVK run (O3b), which ranges between November 1, 2019 and March 27, 2020. We analyse the two-detector network composed by the LIGO detectors \cite{ collaboration2015advanced}  (data are freely accessible at Gravitational Wave Open Science Center \cite{abbott2023open}). The addition of the Virgo detector,  while it enhances both waveform reconstruction and sky localization, it does not increase the search sensitivity due to its lower
sensitivity and unfavourable co-alignment with LIGO detectors \cite{O3allskyshort}. In-depth analyses on the quality of the data can be found in Ref. \cite{davis2021ligo}. The coincident time between the LIGO detectors during O3b is 95 days. We accumulate about 305 years of background, using the time-shift analysis: we employ around 25$\%$ of it to train the XGBoost model and the remaining to assess the significance of the detected triggers.
\section{Results}
In this section we present the results of the analysis on O3b data. We report also the search sensitivity achieved in terms of spacetime volume and distance, and we constrain the events rate. Using the same methodology, we present the sensitivity prospects for the next observing runs with current ground-based detectors.
\subsection{Search}
\begin{figure}[ht]
\includegraphics[scale=0.35]{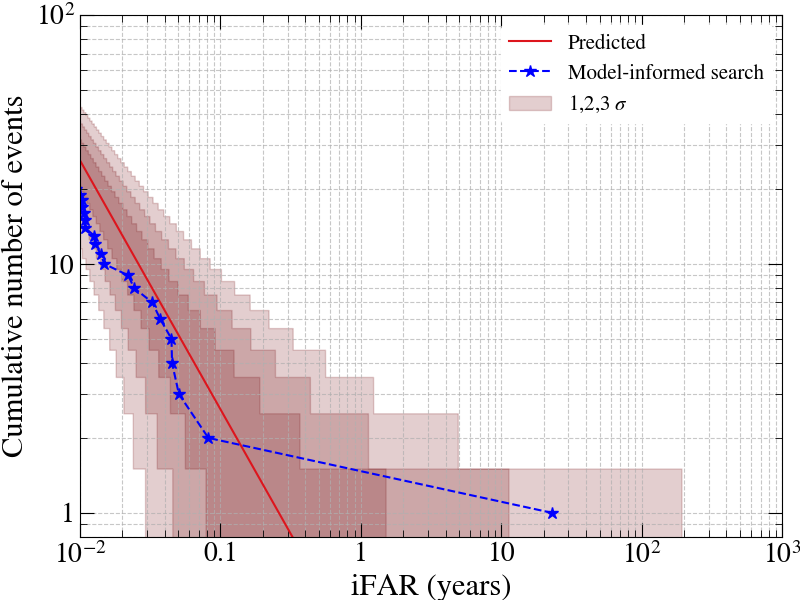}	\caption{Cumulative number of events versus iFAR found by the model-informed search on O3b data. The red solid line shows the expected mean value of the background, 1$\sigma$, 2$\sigma$ and 3$\sigma$ Poisson uncertainty intervals are reported with shaded region. There are no significant outliers. }
	\label{fig:zerolag}
\end{figure}

No significant event has been identified in addition to known detections of CBC \cite{gwtc3}. The results of the search are reported in Fig. \ref{fig:zerolag}. The most significant trigger, at iFAR $\sim$ 20 years has SNR of 8.5 and 8.6 in LIGO Livingston and LIGO Hanford respectively, peak frequency at 89 Hz, bandwidth of 37 Hz and duration of 0.012 s. This trigger has been reported in the O3b LVK catalog \cite{gwtc3} as the CBC event GW191222, detected with an iFAR $> 1100$ years, at a distance of $3.0 \pm 1.7 $ Gpc. It is one of the most massive CBC event observed during O3b: the  component masses are $m_1 = 45.1^{+10.9}_{-8.0} M_{\odot}$ and $m_2 = 34.7^{+9.3}_{-10.5} M_{\odot}$. As it is originated from a massive system, it has a short duration which can explain why it results the most significant trigger in our search. There is no significant evidence of a waveform deviation w.r.t to CBC template \cite{gwtc3}: the waveform consistency test, briefly described in note \footnote{The waveform consistency test \cite{salemi2019wider} computes the fitting factor (Eq. \ref{eq:ff}) \textit{on-source} between the cWB reconstructed waveform and the waveform generated using compact binary coalescence parameter estimate (CBC PE). This value is then compared with a null distribution created injecting CBC PE template near the time of the event, and computing the fitting factor \textit{off-source} between such injections and their cWB reconstruction. If the data period is noisy or the analysis algorithm is particularly biased for such signals, the null-distribution will be broader. For the CBC event GW19122 the fitting factor \textit{on-source} is 0.91 and the fitting factor \textit{off-source} is $0.88^{+0.04}_{-0.07}$, indicating that there is no significant discrepancy, and the CBC model is explaining accurately the data.}, compares the waveform reconstructed with minimal assumption by cWB and the CBC waveform model and indicates that they are in agreement. It is not surprising that the search proposed here does not find significant CBC events: the model-informed search is tuned towards HE signals, whose time-frequency evolution differs from typical CBC signals. 

\subsection{Search sensitivity}
\label{sec:search_sens}
A key aspect of this work is the assessment of the observable sensitivity spacetime volume $\langle V T \rangle$, that represents the portion of the Universe in which the proposed analysis would have detected a HE signal with a certain significance, if any \cite{tiwari2018estimation}.  
In order to compute $\langle V T \rangle$, HE simulations are uniformly generated in a volume $V_0$ defined up to a maximum redshift $z_{\mathrm{max}}$:
\begin{equation}
    V_0 = \int_{0}^{z_{\mathrm{max}}}\frac{\mathrm{d}Vc}{\mathrm{d}z} \frac{1}{1+z} \mathrm{d}z
    \label{eq:V}
\end{equation}
where $\frac{\mathrm{d}Vc}{\mathrm{d}z}$ is the differential co-moving volume and $\frac{1}{1+z}$ accounts for the Universe expansion. $z_{\mathrm{max}}$ is chosen so that the detection efficiency becomes negligible. The average sensitivity spacetime volume is computed from $V_0$ and the recovering efficiency as:
\begin{equation}
    \langle VT \rangle = V_0 \frac{N_{\mathrm{det}}}{N_{\mathrm{inj}}} T
    \label{eq:sens_V}
\end{equation}
where $N_{\mathrm{inj}}$ is the number of signals injected, $N_{\mathrm{det}}$ the number of events detected by cWB with a certain significance, and $T$ the observation time. In the following, we consider \textit{detected} the simulations recovered with iFAR $>$10 years. Different choices of iFAR thresholds are discussed in Appendix \ref{sec:choiceiFAR}. Fig.\ref{fig:distance} reports the sensitive spacetime volume in each mass range: at lower masses, the GW energy released during the encounter is less and so only closer events could be detected. Instead, at high masses the peak frequency decreases, and enters the band where the detectors are less sensitive. The uncertainty on the sensitive spacetime volume estimate arises from the statistical errors due to the finite number of injections and the calibrations errors of the GW detector output. The statistical errors can be estimated considering the binomial statistics which gives $dV_{stat} = \frac{1}{\sqrt{N_{det}}} \sigma(N_{det}/N_{inj})$ \cite{tiwari2018estimation}, and is $<2\%$ for each mass range considered. The major contribution to the uncertainty arises from GW data calibration error which in amplitude is $< 12\%$ \cite{sun2021characterization}, and it translates into an error on  sensitive spacetime volume $< 36\%$. $\langle V T \rangle$ is minimum for HE between compact objects with masses in the range [2, 5] $M_{\odot}$ (0.059$\pm 0.021 \times  10^{5}\mathrm{Mpc}^3$ year) and maximum for masses in [20, 40] $M_{\odot}$ (3.9$\pm 1.4\times 10^{5} \mathrm{Mpc}^3$ year). $\langle V T \rangle$ estimates are reported in Tab.\ref{tab:results}.

In order to compare our results with literature, it is useful to introduce the sensitive distance $D$ which is derived easily from the sensitive volume as:
\begin{equation}
  D = \left( \frac{3 \langle V T \rangle}{4 \pi T_a} \right)^{1/3}  
  \label{eq:sens_D}
\end{equation} 
where $T_a$ is considered equal to one year.
\begin{figure}[t]
\includegraphics[scale=0.5]{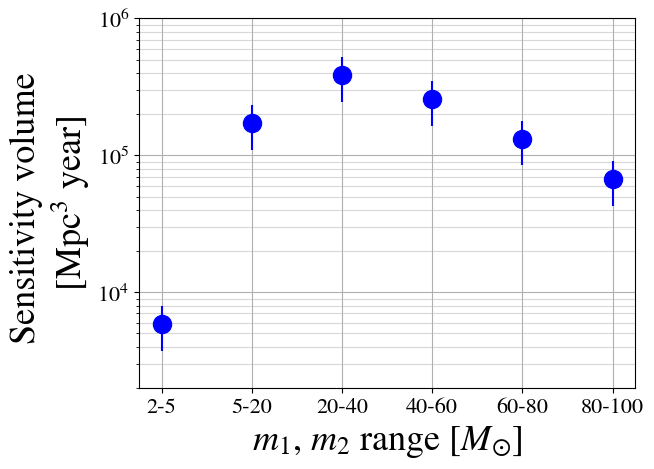}	\caption{Sensitivity spacetime volume $\langle V T \rangle$ for each mass range considering hyperbolic encounters simulations recovered by cWB with an iFAR $>$ 10 years.}
	\label{fig:distance}
\end{figure}
The farthest distance, 45.1$\pm 5.4$ Mpc, is reached when the component mass are between 20 and 40 $M_{\odot}$.

We compare our findings with the corresponding estimates found in Ref. \cite{dandapat2023gravitational}, which are computed using the same waveform family, assuming Advanced LIGO power spectral density (PSD) and a matched filter search. With such configuration, optimally placed HE between two NS and two BH are visible for Advanced LIGO up to $\sim 20$ and $\sim 170$ Mpc, respectively. These estimates are a factor $\sim 2 - 4$ higher with respect to the sensitive distances reported in Table  \ref{tab:results}. The reasons of such differences are twofold: first our search analyses the O3b strain data which has a lower sensitivity than the one considered in Ref. \cite{dandapat2023gravitational}. 
In addition, the prospect analysis considered a match filter search, which typically has a better performance than a model-informed search. As discussed in Section \ref{sec:cwb}, a model-informed search does not completely rely on the accuracy of the waveforms, and it is sensitive to a wider range of morphologies. This advantage implies that the discrimination of GW signals from transient noises is more difficult, and the search is less sensitive w.r.t to the match filter analysis.

As the search does not report any GW signal associated with HE, we report the upper limit of the rate per unit volume at the $90\%$ confidence level, defined as \cite{sutton2009upper}:
\begin{equation}
    R_{90\%} = \frac{2.303 }{ \langle VT \rangle}
\end{equation}
where $2.303 =-\mathrm{log}(1-0.9)$. The uncertainty of the rate is obtained propagating the error on the spacetime volume. The most stringent rate is 5.98$\pm 0.94\mathrm{Mpc}^{-3}\mathrm{Myear}^{-1}$, achieved for HE with masses in [20, 40] $M_{\odot}$. Table \ref{tab:results} summarizes the sensitive spacetime volume, distance and expected rate in each mass range.
\begin{table*}
\begin{centering}
\begin{tabular}{|M{4.5cm} || M{1.8cm} M{1.8cm} M{1.8cm} M{1.8cm} M{1.8cm} M{1.8cm}|} 
 \hline
 Mass range [$M_{\odot}]$ & $[2, 5]$ & $[5, 20]$ & $[20, 40]$ & $[40, 60]$ & $[60, 80]$ & $[80, 100]$ \\
 \hline
 \hline
 Volume $[10^{5}\mathrm{Mpc}^3\mathrm{year}]$&0.059$\pm$0.021&1.71$\pm$0.61&3.9$\pm$1.4&2.58$\pm$0.93&1.32$\pm$0.47& 0.67$\pm$0.24\\ 
 \hline
 Distance [Mpc] & 11.2$\pm 1.3$&34.4 $\pm4.1$ & 45.1$\pm$5.4 &  39.5$\pm 4.7$ & 31.6$\pm3.8$& 25.2$\pm3.0$\\
 \hline
 Rate [$\mathrm{Mpc}^{-3} \mathrm{Myear}^{-1}$] & 392$\pm61$& 13.4$\pm2.1$ &5.98$\pm$0.94 &8.9$\pm$1.4&17.5$\pm$2.7&34.3$\pm$5.4\\ 
 \hline
\end{tabular}
\end{centering}
 \caption{Sensitive spacetime volume, distance and event rate for hyperbolic encounter simulations recovered with an iFAR $>$ 10 years in each mass range. The errors reported are mainly due to GW detector output calibration uncertainties.}
 \label{tab:results}
 \end{table*}
\subsection{Properties of the HE simulation recovered}
\begin{figure*}[t]
\includegraphics[scale=0.5]{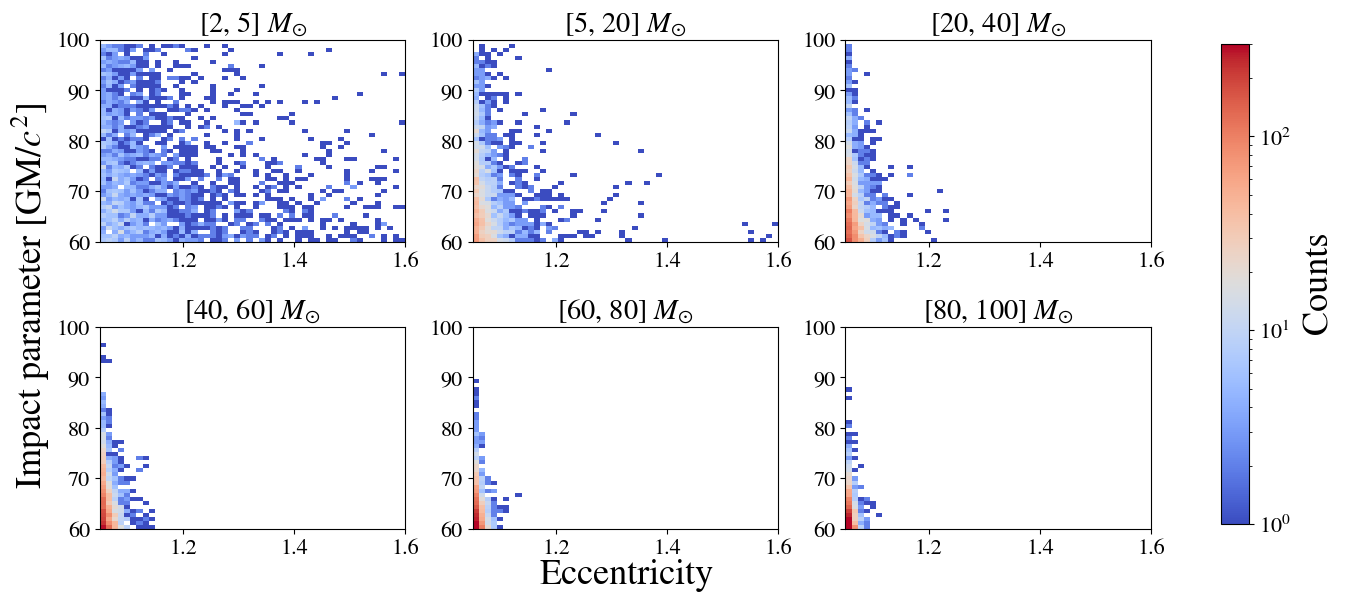}
	\caption{Impact parameter versus eccentricity for the HE injections recovered by cWB with iFAR$>$ 10 years. Each frame refers to a different mass range on the compact objects involved (increasing from left to right). The injections are uniformly distributed in the parameter space represented, but strong selection effects are evident with the increases of the component masses.}\label{fig:b_and_ecc}
\end{figure*}

In this section we discuss the properties of the HE injections recovered by cWB. The impact parameter and the eccentricity are uniformly distributed between [60,100] $GM_{tot}/c^2$ and [1.05,1.6] in the injections sets. However, the detected events by cWB cover only a certain area of such parameter space (Fig. \ref{fig:b_and_ecc}). In particular, with the increase of the total mass only the events with low impact parameter and low eccentricity are recovered. As described in Section 2, HE with high impact parameter and high eccentricity have a lower frequency peak, and when the total mass is high the signal is at the boundary of the frequency range of the detectors, where the sensitivity is worse. Such selection effects have to be taken into account accurately to infer source populations properties, especially in the case of a detection of HE signal.

\begin{figure}[ht]
\includegraphics[scale=0.5]{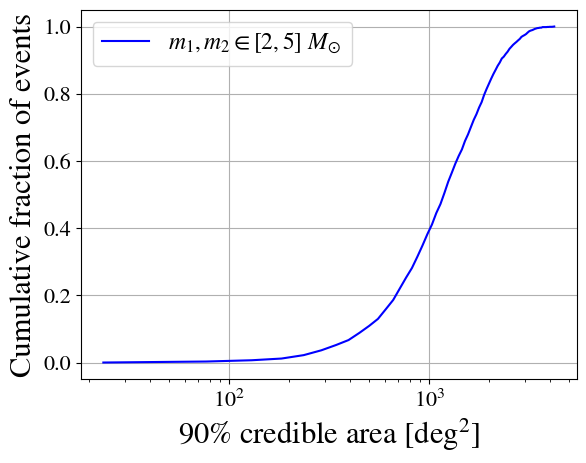}	\caption{Cumulative fractions of events detected with iFAR $>$ 10 years versus 90\% credible area for HE with component masses in the range [2, 5] $M_{\odot}$ using O3b data. About 40\% of the detected simulations are localized with a search area of $\leq 1000 \ \mathrm{deg}^2$.}
\label{fig:search_area}
\end{figure}

Another aspect that is worth discussing is the capability of the search algorithm to localize the source. HE between neutron stars might have an electromagnetic counterpart, as described in Section \ref{sec:astro}. In order to establish an association between the GW signal and a counterpart, the algorithms should localize the source in a small area of the sky. In Fig. \ref{fig:search_area} we report the 90\% credible area for HE with component masses between [2, 5] $M_{\odot}$ obtained with cWB. We show that about $40\%$ of the events detected with iFAR $>$ 10 years have a search area of $\sim 1000$ deg$^2$. Such a poor localization would make difficult multi-messenger follow-up searches. The search area improves significantly including more GW detectors in the analysis \cite{abbott2020prospects}: in the Appendix \ref{sec:appendix_prospect} we report the localization prospects for future observing runs including Virgo \cite{acernese2014advanced} and KAGRA \cite{akutsu2021overview} detectors.

\subsection{Prospect for future observing runs}
\label{sec:prospect}
Due to the continuous upgrade of the detectors, the possibility of detecting GW from close HE becomes more probable as we move onto future observing runs. Here, we estimate the increase in sensitivity expected for LVK observing runs O4 and O5 considering HE in the mass range $[20,40] M_{\odot}$, since this gives our maximum spacetime volume. We consider the three detectors network HLV considering LIGO and Virgo detector, and the four detectors network HLVK including also KAGRA detector.  We simulate Gaussian noise to represent the noise floor from the future detectors, using the expected PSD for O4 and O5 observing runs \cite{abbott2020prospects}. More details on the search configuration in the case of Gaussian noise are reported in Appendix \ref{sec:appendix_prospect}. In order to account for the increased efficiency, the HE waveforms are injected into the simulated data over increased volumes $V_0$. 

\begin{figure}[ht]
    \includegraphics[scale=0.5]{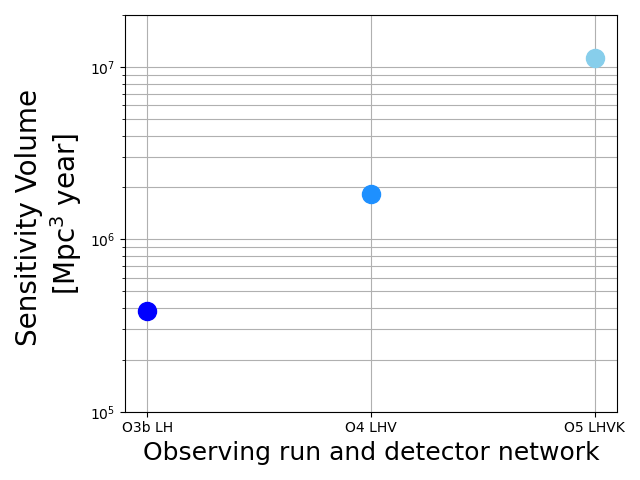}
    \caption{Sensitivity volume estimates of hyperbolic encounters in mass range $[20,40] M_{\odot}$ for future observing runs. A detection threshold of iFAR $>$ 10 years is considered.}
    \label{fig:prospect}
\end{figure}

We estimate our sensitive volume as in Eq \ref{eq:sens_V}, while only considering the binomial statistical errors. Fig. \ref{fig:prospect} shows sensitive volume at iFAR $>$ 10 years in O4 and O5, with the O3b results shown for comparison. For O4, we see that with the 3-detectors LHV network we may be sensitive to HE events up to a volume of 1.830$\pm 0.048 \times 10^6$ Mpc$^{3}$ year, while in O5 the LHVK network may achieve a sensitive volume of 1.133$\pm 0.052 \times 10^7$ Mpc$^{3}$ year. Multiple detector networks significantly improve the localization of GW signals, as shown in Appendix \ref{sec:appendix_prospect}. 

\section{Astrophysical implications}
\label{sec:astro}
Several studies compute the event rate for close HE. In a simplistic derivation, the expected event rate is computed from the individual collision rate $\tau = n v \sigma$, where $n$ is the number density of compact objects, $v$ their relative velocity, and $\sigma$ the cross-section which depends on the impact parameter $b$ as $\sigma \sim \pi b^2$. The detectable rate depends strongly on the interplay between the energy released in GW and the cluster's properties: the first depends mainly on the mass of the sources, the impact parameters $b$ and the eccentricity $e$, while astrophysical considerations fix $n$ and $v$. Ref. \cite{mukherjee2021gravitational}  focuses on HE of compact stars in globular cluster. The expected event rate per year is $9 \times 10^{-10}$Mpc$^{-3}$year which is a factor $\sim 6000$ lower than our most constraining rate.

Ref. \cite{garcia2017gravitational} presents the case of HE between primordial black holes (PBH). The expected rate per volume computed for such sources, assuming initial velocities between PBH of 200km/s, is $\sim$ $0.016 (\frac{b}{25 GM/c^2})^2 (\frac{M_{\mathrm{PBH}}}{30 M_{\odot}})^{-2} \mathrm{Mpc}^{-3}\mathrm{Myear}^{-1}$. Considering $b$ of $80 GM/c^2$ the expected rate is $\sim 0.16 \mathrm{Mpc}^{-3}\mathrm{Myear}^{-1}$, a factor 40 below our best rate.

Hyperbolic encounters involving a NS might be associated with an electromagnetic counterpart, referred to as \textit{shattering flares}   \cite{tsang2013shattering}. 
Briefly, the mechanism of such emission is the following: during a close encounter part of the kinetic energy of the orbit is transferred into resonant tidal coupling. If the energy involved is enough, seismic oscillations of the NS might couple to the magnetic field producing a strong transverse electric field which accelerates particles to high energy. The luminosity of such shattering flares are estimated of the order of $\sim 10^{47}-10^{48}$erg/s in X-ray and soft gamma-ray bands.
Such events are rare, but the possibility of observing a multi-messenger event make them very interesting: an association with electromagnetic counterpart would help to distinguish the GW event from glitches, facilitating the event validation, and would provide incredible insight into the interaction, especially on constraining the equation of state of the NS. In order to associate a GW event with a counterpart, the localization is crucial. cWB localization capabilities are discussed in Fig.\ref{fig:search_area} for O3b data and in Fig. \ref{fig:search_area_prospect} for future observing runs. The expected rate of HE involving NS is of the order of $3\times 10^{-6} \mathrm{Mpc}^{-3}\mathrm{Myear}^{-1}$, which is 8 orders of magnitude from our rate for component masses in [2, 5] $M_{\odot}$ \cite{tsang2013shattering}.

The prospects presented in Section \ref{sec:prospect} show that for the O4 observing run the sensitivity volume will increase by a factor $\sim 5$, and for O5 by a factor $\sim 27$ w.r.t to O3b data. 
According to current estimates in case of null detection of HE signal, we would be able to constraint extreme clusters models, and we might discuss PBH density in the next observing runs. Instead, HE between neutron stars have an expected rate far below current detector capabilities.

\section{Conclusions}
Hyperbolic encounter (HE) between stellar mass compact object are expected to emit GW burst in the frequency of current ground-based detector. We search for HE signals in O3b  LVK data with the weakly-modeled algorithm Coherent WaveBurst. We employ the decision tree algorithm XGBoost trained with HE signals to build a model-informed search. One of the major challenges when searching for such source is the similarity between HE waveforms and \textit{blips}, one of the most common short-duration noises present in GW data. To mitigate the impact of such disturbances, we include in the cWB algorithm an autoencoder neural network which identifies blip glitches and improves their rejection.

No significant event has been identified in addition to known detections of compact binary coalescences. We inject 3 PN-accurate waveforms in real data varying the  component masses between [2, 100]$M_{\odot}$, the impact parameter in [60, 100] ${GM}/{c^2}$ and the eccentricity in [1.05, 1.6]. For the first time to our knowledge, we compute sensitivity volume for such sources: the best reach is 3.9$\pm 1.4 \times 10^5$Mpc$^3$year achieved for HE with masses between 20 and 40 $M_{\odot}$, while the less constraint range is $0.059\pm0.021 \times 10^5$Mpc$^3$year obtained for masses between 2 and 5 $M_{\odot}$. We also estimate the sensitivity prospects for HE events in future LVK observing runs: the sensitivity volume in O4 considering LIGO and Virgo detectors will be  $1.83 \times 10^6 \mathrm{Mpc}^3$year, and for O5 including also KAGRA detector the sensitivity will reach $1.13 \times 10^{7}\mathrm{Mpc}^3$year.
We also provide upper limits of the rate at 90\% confidence level, and we compare our findings with the expected rates given in the literature.

The prospects for future observing run are encouraging for pursuing the searches of HE. Unless a significant decrease in the transient noises rate, such disturbances represent the limiting factor of the search sensitivity, and would challenge a candidate event validation in case of a significant event. In future works, we plan to face this issue exploiting also GW polarization that would be distinguishable with the detector networks of the future observing runs.
\begin{acknowledgments}
We would like to thank Gonzalo Morras for their comments on the manuscript. This material is based upon work supported by NSF's LIGO Laboratory which is a major facility fully funded by the National Science Foundation. The authors are grateful for computational resources provided by the LIGO Laboratory and supported by National Science Foundation Grants PHY-0757058 and PHY-0823459. This research has made use of data, software and/or web tools obtained from the Gravitational Wave Open Science Center, a service of LIGO Laboratory, the LIGO Scientific Collaboration and the Virgo Collaboration. S. B. is grateful for the support of the Pauli Center for Theoretical Physics and the University of Zurich, Switzerland. S.T. is supported by Swiss National Science Foundation (SNSF) Ambizione Grant Number : PZ00P2 - 202204. Y.X. is supported by China Scholarship Council. G.P. acknowledges support by ICSC – Centro Nazionale di Ricerca in High Performance Computing, Big Data and Quantum Computing, funded by European Union – NextGenerationEU.
\end{acknowledgments}

\appendix
\section{Choice of iFAR threshold}
\label{sec:choiceiFAR}
\begin{figure}[ht]
\includegraphics[scale=0.26]{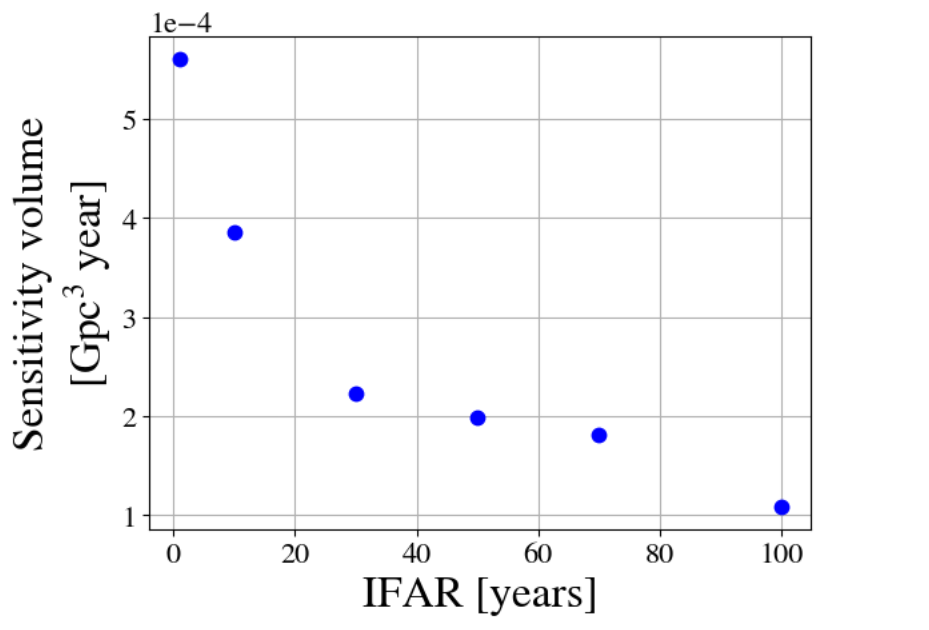}	\caption{Sensitive volume at different iFAR thresholds for HE with masses in [20, 40] $M_{\odot}$. The results reported in Section \ref{sec:search_sens} consider as threshold iFAR $>$ 10 years.}
\label{fig:volume_ifar}
\end{figure}

In Section \ref{sec:search_sens} and \ref{sec:prospect}, we report the sensitivity volume, distance and event rate considering HE simulations detected with iFAR $>$ 10 years. Here, we discuss the choice of the iFAR threshold. Fig.\ref{fig:volume_ifar} shows how the sensitivity spacetime volume (Eq. \ref{eq:V}) varies with different iFAR thresholds for HE with masses in the range [20, 40] $M_{\odot}$. A similar trend is obtained for the other mass ranges. Higher iFAR thresholds constrain more the search sensitivity, but are more subjected to statistical fluctuations: the number of detected simulations decreases with higher iFAR threshold and strongly depends on the high SNR tail of the background distribution, which is populated by loud short-duration noises. For this reason, we consider the choice of iFAR$>$10 years suitable. 

\section{The case of gaussian noise and future localization capabilities}
\label{sec:appendix_prospect}
The sensitivity prospects presented in Section \ref{sec:prospect} are achieved injecting HE waveforms on the  PSD expected for the future LVK observing runs. The noise is assumed stationary and gaussian, meaning that there are no glitches. Hence, we analyse these datasets using cWB algorithm without applying the XGBoost signal-noise separation procedure (Section \ref{sec:cwb}). A natural question to understand the volume sensitivity prospects is how different is the efficiency of our search on gaussian data w.r.t real data.
In order to compare the two configurations, we perform a set of HE injections on O3b PSD gaussian data: we find that considering a detection threshold of iFAR $>$10 years the search sensitivity on gaussian data is similar w.r.t to the real case scenario. Using this threshold, the XGBoost model is able to efficiently reduce the background distribution. The differences emerge at higher iFAR where the gaussian noise background distribution disappears, while in the real data case there are high SNR glitches whose mitigation is particularly difficult. Hence, we can conclude that the prospects estimates given in Section \ref{sec:prospect} do not depend on the gaussian noise hypothesis.

\begin{figure}[ht]
\includegraphics[scale=0.5]{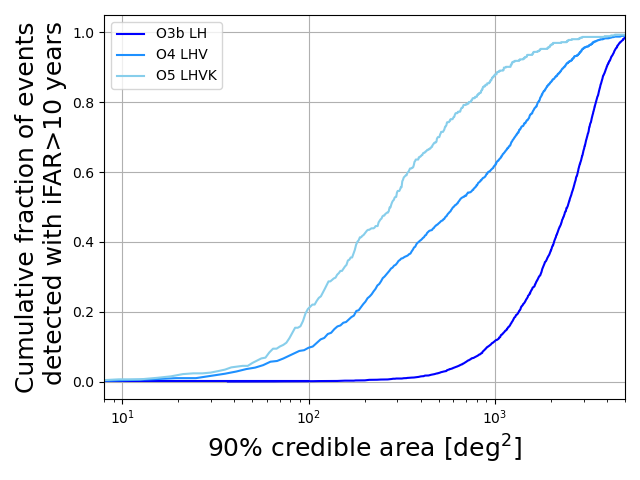}	\caption{Cumulative fractions of events detected with iFAR $>$ 10 years versus 90\% credible area for HE with component masses between [20, 40] $M_{\odot}$. We compare HL O3b results (dark blue) with HLV network in O4 (blue) and HLVK network in O5 (light blue).}
\label{fig:search_area_prospect}
\end{figure}

We also present the localization prospect for future observing runs, similarly to the result shown for O3b data in Fig. \ref{fig:search_area}. The more the GW detectors, the better is the localization of the GW signals \cite{abbott2020prospects}. We show cWB 90\% credible area for HE also in HLV network using O4 PSD, and HLVK network using O5 PSD in Fig. \ref{fig:search_area_prospect}. Multiple detectors networks significantly improve the localization: in the mass range between [20, 40] $M_{\odot}$ about 16\% of the detected simulations are localized with a search area of $\sim$ 1000 deg$^2$ for HL network, $\sim$ 60\% for HLV in O4 and $\sim$ 86\% for HLVK in O5.

\bibliography{apssamp}

\end{document}